\documentstyle[psfig,epsf]{aipproc}
\input psfig.tex
\begin{document}
\title{Jets and Outflows From Advective Accretion Disks}

\author{Sandip K. Chakrabarti\thanks{Also, Honorary Scientist, Centre for Space Physics, IA-212, Salt Lake, Calcutta 700097}}
\address{S.N. Bose National Centre for Basic Sciences\\
JD Block, Salt Lake, Calcutta 700098\\}

\maketitle

\begin{abstract}
Jets and outflows must be produced directly from accretion disks and inflows, 
especially when the central gravitating objects are compact, such as neutron stars
and black holes, and themselves are {\it not} mass losing. Here, we review the 
formation of jets from advective inflows. We show that the centrifugal
pressure supported boundary layer (CENBOL) of the black holes may play 
crucial role in producing outflows. CENBOL is not present in Keplerian
disks. Thus energetic jet formation is directly connected to sub-Keplerian
flows close to compact objects.
\end{abstract}

\noindent To appear in Proceedings of Heidelberg International Symposium on Gamma-ray
Astronomy (AIP publication). F. Aharonian and H. Voelk (Eds.)

\section*{Introduction}

Jets and outflows are ubiquitous in quasars and active galaxies.
They are often superluminal, moving almost at the speed of light.
Over the last decade, a few sources in our own galaxy have been observed
which also display energetic outflows. Most interesting of these sources
is the black hole candidate GRS1915+106~\cite{ct92,mr94}
which has been studied very extensively and in 
this review we shall devote some space on this object. 

Extensive work has been carried out over last quarter of a century 
to explain the origin of jets~\cite{mr99}. Not surprisingly, most of these invoke
accretion flows to be the source of outflowing matter. Since the standard Keplerian
accretion disk by Shakura \& Synyaev~\cite{ss73}
did not have any scope of the formation of jets, in the early 80's,
accretion solutions of purely rotating disk were improved 
to include the effect of radiation pressure~\cite{mrt76,pw80,lt80,pb81,mp82} 
(See  Chakrabarti~\cite{c96a} for a review.) and see if these thick disks were useful for the
production and collimation 
of jets. Jet solutions have changed from speculative ideas such as 
de-Laval nozzles~\cite{br74} to electrodynamically acceleration model~\cite{z78}, self-similar 
centrifugally driven outflows~\cite{bp81}, `cauldrons'~\cite{br84} etc. Centrifugally 
driven outflows are subsequently modified to include accretion disks~\cite{k89}.
Chakrabarti \& Bhaskaran~\cite{cb92} (see also, Contopoulos~\cite{c95})
showed that it is easier produce outflows from a sub-Keplerian inflow.

It is important to note that in black hole accretion, matter is relativistic
close to the black hole horizon, while in jets, matter become relativistic
farther away from the the black hole. In both the cases, matter is assumed to begin its
journey subsonically and end its journey supersonically. Bondi~\cite{b52}
studied simplest form of accretion and wind solutions with these properties
for spherically symmetric flows without any heating and cooling. Especially important are the 
outflow solutions which were immediately used to study solar and stellar winds~\cite{ha70,enp79}.
In the context of Active Galaxies and Quasars,
some notional unification of disk-like and jet-like solutions were presented by
Chakrabarti~\cite{skc84,skc85}. It was shown that the same solution of purely rotating flow around a 
black hole could describe accretion flows on the equatorial 
plane and pre-jet matters near the axis. With a natural 
angular momentum distribution of $l(r) = c \lambda (r)^n$, 
(where $c$ and $n$ are constants and $\lambda$ is the von
Zeipel parameter) it was found that for large $c$ and 
small $n$ ($n<1$), solutions are regular on the equatorial 
plane and they describe thick accretion disks. However, 
for small $c$ and large $n$ ($n>1$), the solutions
are regular on the polar axis and they describe pre-jet matter.  
It was conjectured that that some viscous process might be 
responsible to change the pair of parameters ($c,n$)
from one set to the other. Eggum, Coroniti \& Katz~\cite{eck85}
considered radiatively driven outflows emerging from a 
Keplerian disk. The angular momentum distribution
in the outflow qualitatively looks similar to the natural 
distribution described above. Chakrabarti~\cite{skc86} used the 
natural distribution to study acceleration of the pre-jet matter
inside the funnel of a thick disk
and found that jets could be accelerated at the cost of 
thermal and rotational energies close to a black hole. Addition of magnetic field
in the wind solution enhances the terminal velocity substantially~\cite{ttaf}.

\section*{Emergence of Global Inflow Outflow Solutions (GIOS)}

\begin{figure}
\vbox{
\vskip -3.5cm
\hskip -0.0cm
\centerline{
\psfig{figure=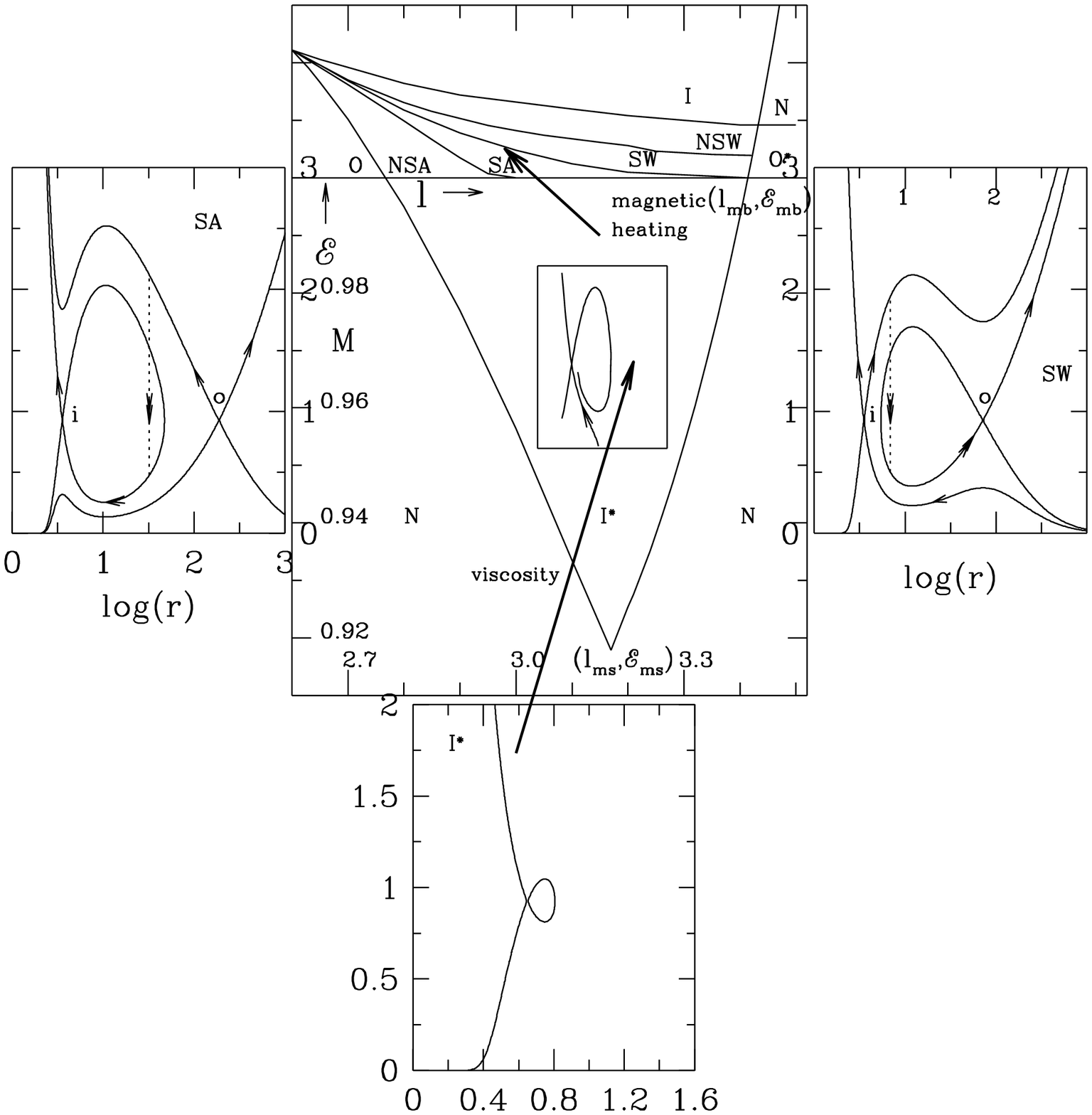,height=10truecm,width=10truecm}}}
\begin{verse}
\vspace{0.0cm}
\noindent{\small{\bf Fig.1:}  Classification of parameter space (spanned by
specific energy and  angular momentum) in several regions depending on whether or not 
solutions include (SA and SW) steady or oscillating (NSA, NSW) shock waves. Solutions from $I^*$ 
region would not be complete unless viscosity is added. These would be regular Keplerian
disks which pass through inner sonic points [inset]. Magnetic heating changes Keplerian
flows into flows with shocks as indicated by dark arrows.}
\end{verse}
\end{figure}

\begin{figure}
\vbox{
\vskip -1.0cm
\hskip 2.0cm
\psfig{figure=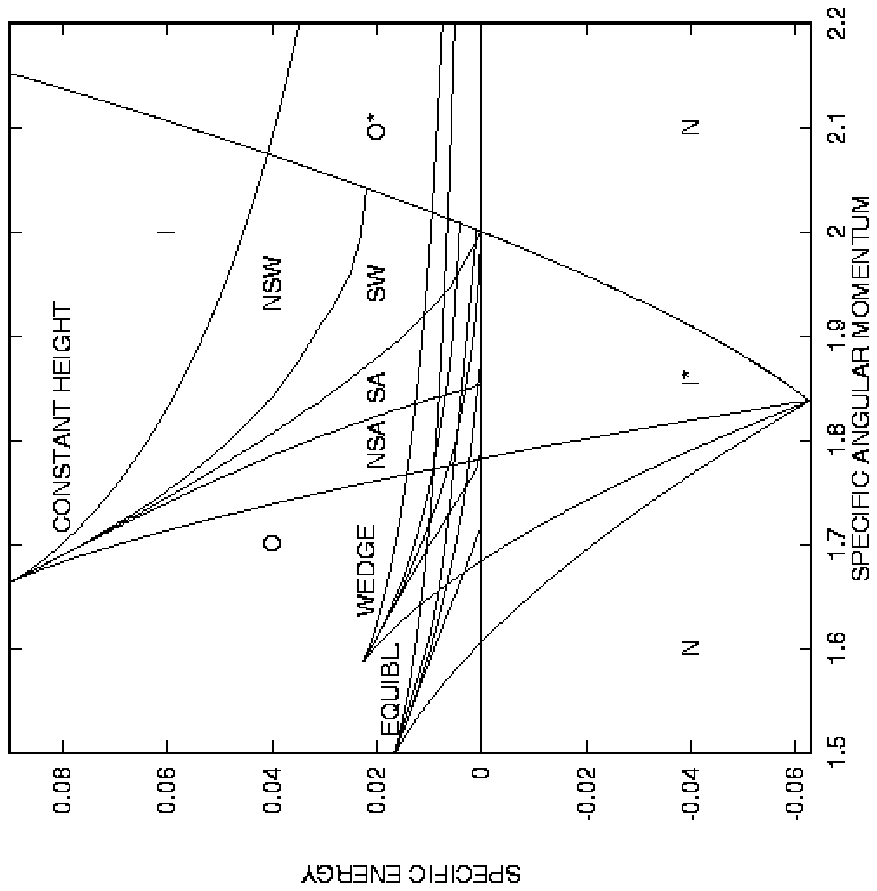,height=7truecm,width=10truecm,angle=90}}
\begin{verse}
\vspace{0.0cm}
\noindent{\small{\bf Fig.2:} Demonstration of the {\it generic} nature of the classification of the
parameter space around a Schwarzschild black hole. Three models (a) a flow which is in
vertical equilibrium as each point (EQUILIB.) (b) a wedge shaped conical flow (WEDGE) and
(c) a flow with a constant height (CONSTANT HEIGHT) are studied. Meaning of various symbols are as in
Fig. 1. Pseudo-Newtonian potential~\cite{pw80} has been used.}
\end{verse}
\end{figure}

Generalized Bondi flows have been solved when angular momentum,
heating and cooling are added~\cite{c89,ttaf,gut}.
All possible accretion and wind type solutions are found, including solutions which
may contain standing shocks and the entire parameter space is classified
according to the nature of solutions~\cite{c89,ttaf}. Fig. 1 shows some of the interesting solutions 
(Mach number is plotted  against logarithmic radial distance [in Units of the
Schwarzschild Radius] in outer three small boxes.)
and the classification of the parameter space
(Specific energy is plotted against specific angular momentum in the central box.)
when the Kerr parameter $a=0.5$ and when the equatorial plane solutions are considered.
The inward pointing arrows in Fig. 1 indicate accretion solutions
and the outward pointing arrows indicate wind solutions. A flow 
from regions I and O passes through the inner sonic point and outer sonic 
point respectively. Those from NSA have no steady shocks in accretion, from 
SA have shocks in accretion, from NSW have no steady shocks in winds, 
and SW have shocks in winds respectively. 
The horizontal line at unit energy in the central box 
represents the rest mass of the inflow. Note that the outflows are produced only when the
specific energy is higher than the rest mass energy. A flow with a lesser energy
produces solutions with closed topologies (I* and O*). 
When viscosity is added the nature of the solutions is changed fundamentally~\cite{skc90}
(see, Chakrabarti~\cite{gut,oebu98} for details) allowing matter to directly come out of a Keplerian
disk and enter into a black hole through the sonic point. The box drawn with I$^*$ region
shows how an I$^*$ solution is modified in presence of viscosity (thick arrow).
When magnetic heating is added or the flow is away from the equatorial plane,
energy of the inflow could become positive and the solutions such as
NSA, SA, SW or NSW could be chosen depending on the parameter space.

This classification is generic and is model independent though the actual 
parameter boundaries would vary. Fig. 2 compares the classification of
parameters in conical flow, the flow in vertical equilibrium
and the flow with constant height in a pseudo-Newtonian geometry~\cite{pw80}.
All the models show the same sub-divisions in the parameter space.

\begin{figure}
\vbox{
\vskip 0.0cm
\hskip 2.0cm
\centerline{
\psfig{figure=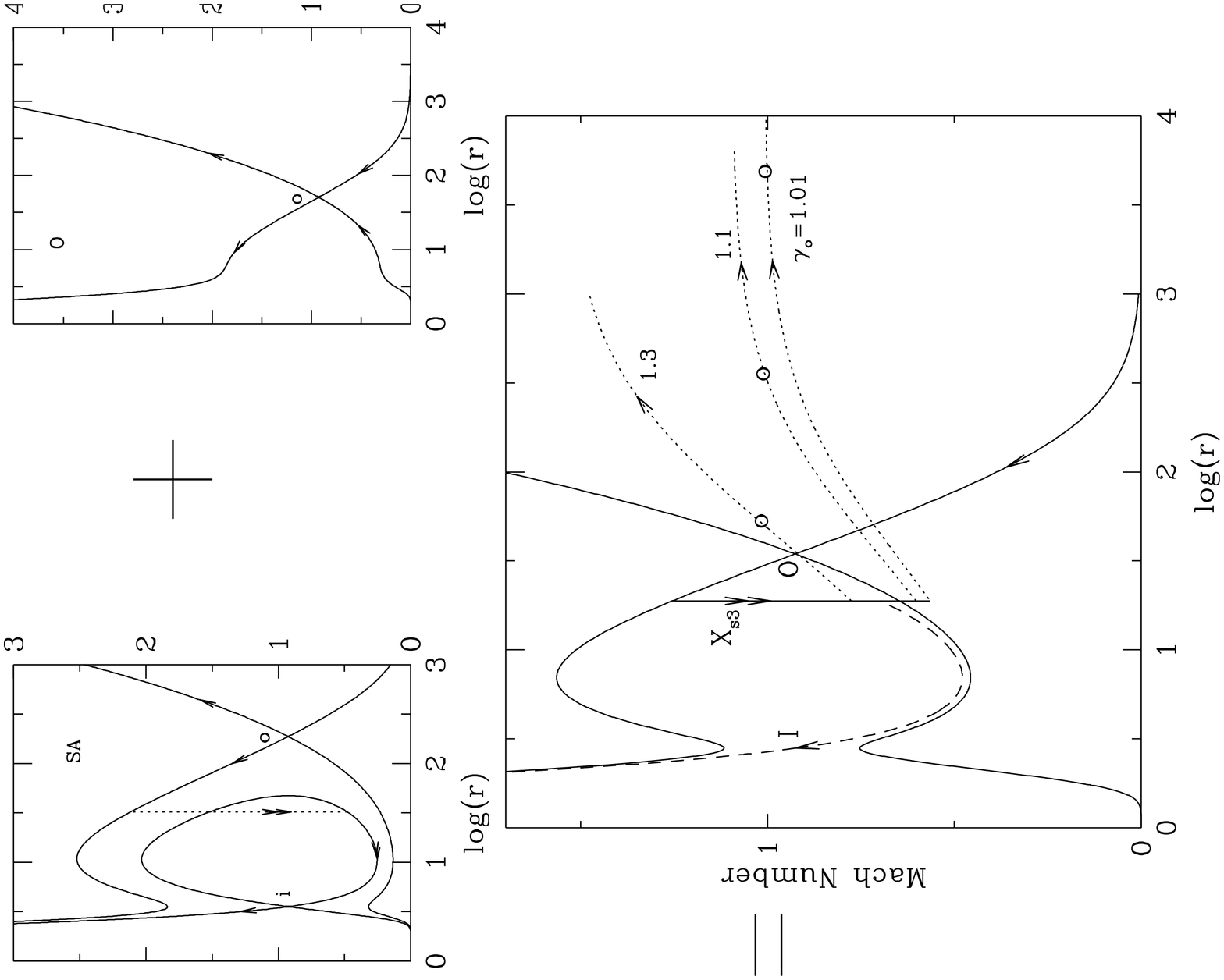,height=7truecm,width=10truecm,angle=90}}}
\begin{verse}
\vspace{0.0cm}
\noindent{\small{\bf Fig.3:} Construction of Global Inflow-Outflow Solutions (GIOS) using 
one solution from SA and one from O regions. Shock formed at $X_{s3}$ generates higher
entropy flow and deflects part of the matter along the axis~\cite{dc99}. }
\end{verse}
\end{figure}

Using these solutions it is possible to self-consistently construct
Global Inflow Outflow Solutions (GIOS) in which matter first accrete using 
solutions with inward pointing arrows and the goes out to long distances
using solutions with outward pointing arrows~\cite{bangla,ijp98a,dc99,caa99}).
Fig. 3 shows how an inflow solution from region SA and an outflow solution from region O 
could be combined to have a GIOS. It has been observed that depending on the 
parameters, the disk could be completely evacuated, by the outflow processes,
especially when the inflow rate is very low~\cite{dc99}.

The steady accretion and wind solutions described above have been verified 
by complete time-dependent simulations~\cite{cm93,mlc94,cm95,mrc96}. 
In the case of steady state solutions, outflows are found to occur 
between the centrifugal barrier and the funnel wall while in 
a non-steady and viscous solution the outflow could spread in regions 
outside the centrifugal barrier as well. Presumably, azimuthal component of the
magnetic field carried along by the outflow is responsible 
for the collimation that is observed in jets.

\begin{figure}
\vbox{
\vskip 0.0cm
\hskip 0.0cm
\centerline{
\psfig{figure=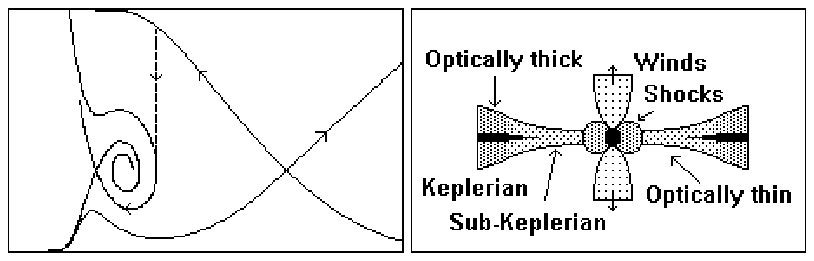,height=8truecm,width=8truecm,angle=90}}}
\begin{verse}
\vspace{-5.0cm}
\noindent{\small{\bf Fig.4:} Topology of SA type solutions in presence of weak viscosity. Shocks
formed (left panel vertical arrow) in accretion increases entropy and the flow is deflected along the 
vertical axis (right panel). Keplerian disk is formed along the equatorial plane where energy is minimum 
and viscosity is the highest.}
\end{verse}
\end{figure}

Fig. 4 shows how the solutions of SA type are modified in presence of 
viscous processed~\cite{ttaf}. For weak viscosity the centrifugal barrier
still remains and shocks form. The post-shock region gets heated and is consequently
puffed up. Winds and outflows are profuse from this region. This is schematically shown
in the right panel of the Figure. On the equatorial plane of the disk, viscosity 
may be high and the disk remains Keplerian. Subsequently, it becomes sub-Keplerian
before entering the black hole. However, the sub-Keplerian flow above and below 
the equatorial plane could be heated up in the corona region by magnetic
dissipation. When its energy become positive (unbound), 
it becomes energetically favorable to produce shocks and outflows.

\section*{Oscillating Shocks in Accretion and Winds}

Solutions in SA and SW are in general steady (with shocks in accretion and 
winds respectively), except when cooling processes are added and cooling time 
scales roughly agree with the infall time scale. Molteni, Sponholz and
Chakrabarti~\cite{msc96} produced examples of this types of shocks. Shocks are
found to perform large amplitude oscillation with a periodicity of
$$
t_{Osc} \sim \frac{R R_s^\alpha R_g}{c \vartheta_0} \ \ \  {\rm s}
\eqno{(1)}
$$
where, $R_s$ is the mean location of the shock in dimensionless units,
$R_g=2GM/c^2$ is the Schwarzschild radius of the central black hole,
$R$ is the compression ratio of the shock, $\alpha\sim 1$ in the
post-shock region. Inverse of these time scales agree very well
with the quasi-periodic oscillations (QPO) of hard X-rays and are believed to be the
cause of QPOs in black hole and neutron star candidates. 
Since cooling time scale generally goes down as accretion rate is increased,
oscillation frequency should go up with accretion rate.

Solutions from the regions NSA and NSW have two saddle type sonic points
as in solutions of SA and SW regions. However, Rankine-Hugoniot conditions
are not satisfied and therefore steady shocks do not form. It was shown that
solutions from these regions still possess shocks, but they are oscillatory,
constantly trying to search for the steady location~\cite{rcm97}. These shock oscillations
also have similar time scales as above, but the oscillation frequency is 
directly sensitive to the angular momentum and specific energy of the flow
and not the accretion rate. Detailed simulation with radiative transfer is
essential to understand these oscillations.

\section*{Rate of outflow generation}

Although hydrodynamic processes alone are not thought to be the
sole mechanisms in generation and acceleration of cosmic jets 
and outflows, a great deal of the behavior of jets could be 
learned about the jets from hydrodynamic considerations. Chakrabarti~\cite{caa99} 
computed the rate of outflow generation assuming the outflow 
to be isothermal at least upto the sonic point. It was found that 
the ratio of the inflow and outflow rates $R_{\dot m}$ could be expressed in terms of 
just three parameters: (a) $R$, the compression ratio of the flow at the
accretion shock, (b) $\Theta_{in}$, the solid angle subtended by the 
accretion flow at the centre and (c), $\Theta_{out}$, the solid angle
subtended by the outflow at the centre. The result~\cite{ijp98a}
is given by:
$$
R_{\dot m}=\frac{{\dot M}_{out}}{{\dot M}_{in}}=
\frac{\Theta_{out}}{\Theta_{in}}\frac{R}{4} 
[\frac{R^2}{R-1}]^{3/2} exp  (\frac{3}{2} - \frac{R^2}{R-1}).
\eqno{(2)}
$$ 
This function $R_{\dot m}$ has a peak when $R\sim 2.5$, i.e., when the shock
is of intermediate strength. In the soft states of black holes, shocks are weakened
as the post-shock region is cooled down completely~\cite{ct95}.
As the compression ratio approaches unity, the outflow rate also goes to zero.
In the hard state, as the compression ratio approaches  $4-7$, $R_{\dot m} \rightarrow 0.1$,
i.e., roughly ten percent of infalling matter goes out of the system.
Details about the relation between the spectral states and the outflows would be 
dealt with elsewhere~\cite{hdpos}.

\section*{Effects of radiative transfer in presence of winds}

Chakrabarti \& Titarchuk~\cite{ct95} showed that post-shock region behaves as the so-called
Compton cloud in determining the spectral states of black hole. In general, when the
soft photons (generated by the Keplerian matter) intercepted by the CENBOL is
very large, this region is cooled down and the black hole is seen in a soft state. 
However, when soft-photons are very few, CENBOL remains hot and the 
black hole spectrum shows characteristics of a hard state. 
Thus state change is equivalent to redistribution of
matter into the Keplerian (producer of soft photons in the pre-shock flow, and hot electrons
in the post-shock region) and the sub-Keplerian components (producers of hot electrons)
in this model. Fig. 5a shows the general disk-jet system in the hard state.

\begin{figure}
\vbox{
\vskip 0.0cm
\hskip 2.0cm
\centerline{
\psfig{figure=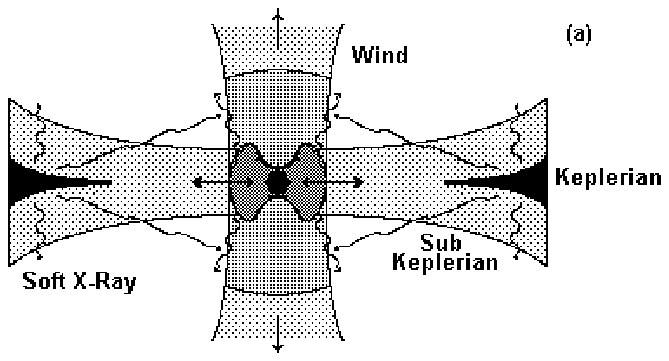,height=7truecm,width=10truecm,angle=90}}}
\vbox{
\vskip -2.0cm
\hskip 2.0cm
\centerline{
\psfig{figure=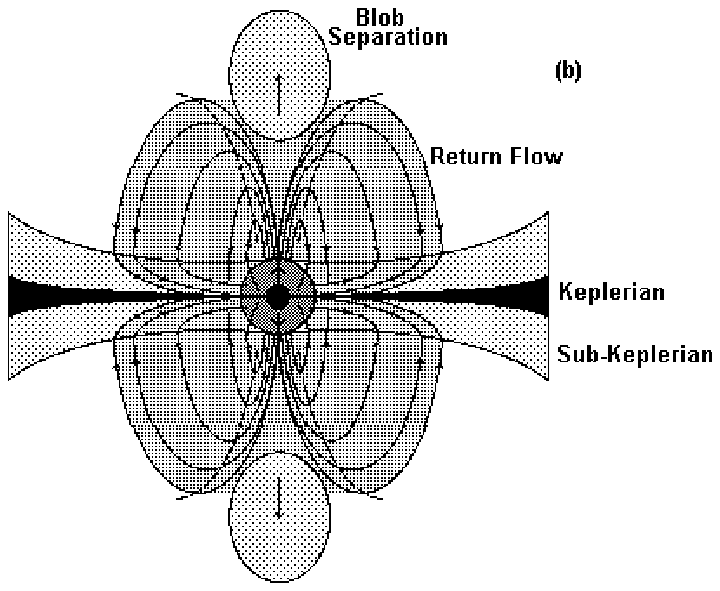,height=7truecm,width=10truecm,angle=90}}}
\begin{verse}
\vspace{0.0cm}
\noindent{\small{\bf Fig.5:} Nature of the accretion and wind solutions in hard states
(a) and in flare states when the sonic sphere is cooled by soft photons from the Keplerian disk (b).
Part of the cooled flow separates as blobby jets and while the rest returns back to the 
accretion disk increasing the accretion rate temporarily.}
\end{verse}
\end{figure}

\begin{figure}
\vbox{
\vskip 0.0cm
\hskip 2.0cm
\centerline{
\psfig{figure=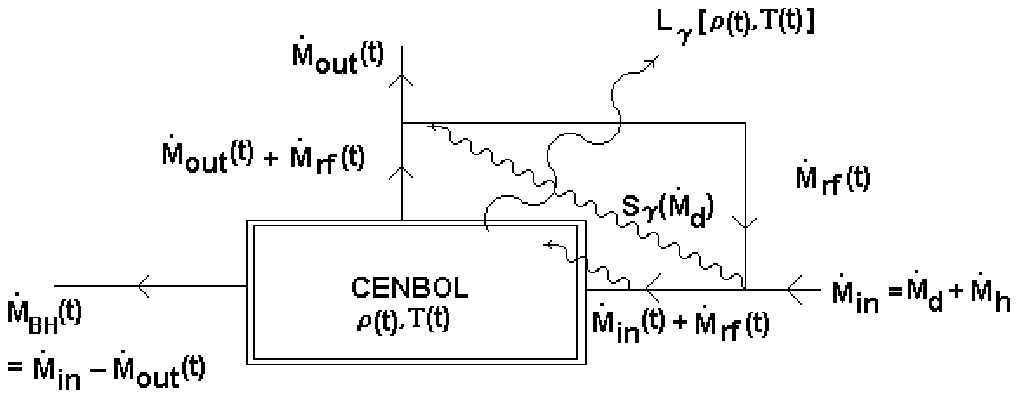,height=7truecm,width=10truecm,angle=90}}}
\begin{verse}
\vspace{-2.0cm}
\noindent{\small{\bf Fig.6:} 
Schematic diagram suggesting the cause of non-linearity in the $L_\gamma $ (t) the light curve.
}
\end{verse}
\end{figure}

There are two very major effects when winds are generated from the CENBOL. In presence of 
significant winds, density of hot electrons in CENBOL goes down, even though the number of 
intercepted soft photons from the Keplerian disk remains the same. The resultant spectrum 
is thus softened~\cite{ijp98b}. When the rate of wind production is very high, 
the region up to the sonic surface of the wind could be cooled down by Comptonization 
process. Since the sound speed in this region is suddenly reduced, the sonic surface of the
wind is shifted downward, and a part of the wind, originally sub-sonic, becomes
supersonic and is separated out as a blob, while the rest below the 
new sonic surface returns back to the accretion disk (Fig. 5b). This causes
enhanced accretion for a brief period and count rates go up~\cite{cm00,cetal00,n00}. 

Figure 6 shows the cause of non-linearity in the light curve $L_\gamma$(t). ${\dot M}_{in}$
is the net inflow  rate which is the sum of the Keplerian rate ${\dot M}_d$ and the
sub-Keplerian rate or halo rate ${\dot M}_h$~\cite{ct95}. This matter together with {\it time
dependent} return flow from the sonic sphere above CENBOL enters CENBOL whose 
density, temperature determines intensity of emitted radiation $L_\gamma$
as well as the outflow rate ${\dot M}_{out}+{\dot M}_{rf}$. CENBOL may emit bremsstrahlung
or Comptonized soft photons. However, $S_\gamma$ the soft photon intensity  intercepted from the
Keplerian disk depends on CENBOL parameters themselves. As Molteni, Sponholz and Chakrabarti~\cite{msc96}
showed CENBOL undergoes oscillations and intercepts photons of varying degree.
The net outflow is ${\dot M}_{out}$ which is also  time-dependent. The net inflow onto the
black hole is ${\dot M}_{BH} = {\dot M}_{in} - {\dot M}_{out}$.

Since the return flow increases electron density in the CENBOL temporarily,
the spectrum is hardened in the high-count states or On states.  
As this excess matter is drained into the black hole, the light curves shows a 
low-count or Off state. The whole process can repeat again and again, though because of 
details of non-linear feedback from the wind results need not be strictly periodic. This
is thought to produce a very interesting behavior in the light curve of 
at least one black hole candidate: GRS1915+105. If the accretion rate is not too high,
the wind rate will be low for most of shock parameters, and such periodic cooling
would not occur. This may be the reason, why other black holes which are
known to have lower rates, do not exhibit as exotic light curve as GRS1915+105.

We thus notice that the spectrum in otherwise harder 
states are softened and softer states are hardened due to the presence of the wind and the
return flows respectively . One
bye-product of this is that the intersection point or the pivotal point of the
two spectra (of On and Off states) recedes to a much higher energy. This is precisely what
is observed in the superluminal source GRS1915+105.
Fig. 7 presents~\cite{cetal00} spectra of three separate days of 
observations and compares them with the low and high
spectra of 1997 March 26 (PID No. 20402-01-21-00)  and of 1997 August 19 with 
(PID No. 20402-01-41-00)~\cite{muno99} respectively. The PID of the three RXTE 
observations are (a) 1997 June 18 (PID 20402-01-33-00), (b) 1997 July 10 (PID 20402-01-36-00)
and (c) 1997 July 12 (PID 20402-01-37-01) respectively. It is very clear that in all these days,
the pivotal point is much farther out compared to the the pivotal points created by Low and 
High states. This is a clear evidence that significant winds are present in 
Off states and a significant return flow is present in the On states.

In a recent paper, Dhawan et al.~\cite{d00} presented a direct correlation between the RXTE-ASM
X-ray data and the IR/Radio observations. If the CENBOL activity (activity which
perturbs Comptonized photons) propagates along the jet
and causes the Radio activity, then the radio activity at around $500$AU
made on 31st October, 1997 would  be perturbed by CENBOL activity
of around 28.5 October, 1997. This assumes that the
perturbation propagates with $0.98c$, the observed speed of the jet. However,
no PCA data is available for 28th or 29th of October. Chakrabarti et al~\cite{cetal00}
compares the spectra of 30th October, 1997 (PID 20402-01-52-02) with the low state data of 
25th of October, 1997 (PID 20402-01-52-00). Here, too a clear evidence of wind was seen.

\begin {figure}
\vbox{
\vskip -9.0cm
\hskip 0.0cm
\centerline{
\psfig{figure=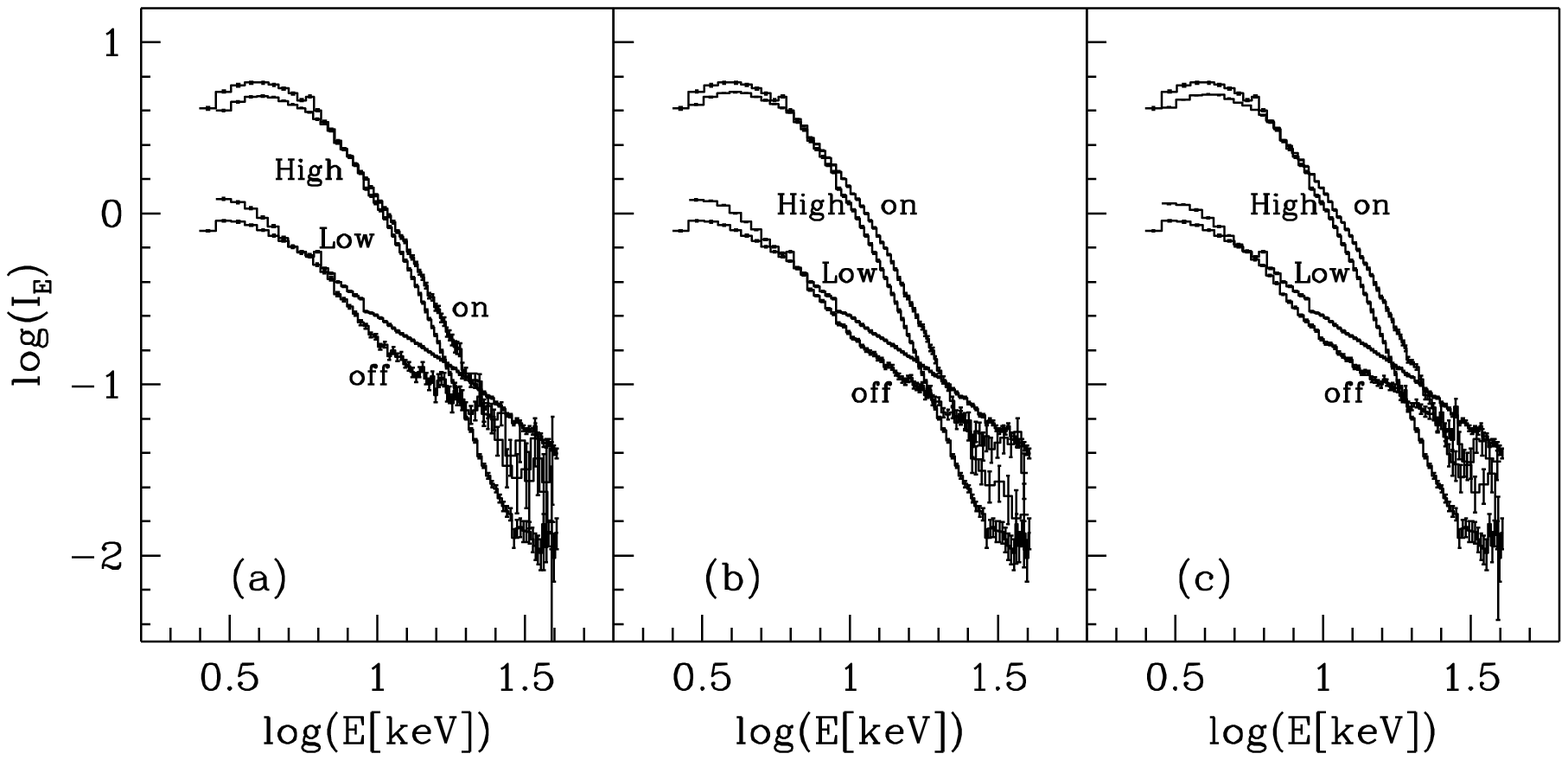,height=15truecm,width=10truecm}}}
\begin{verse}
\vspace{0.0cm}
\noindent{\small{\bf Fig. 7:} 
Unfolded RXTE-PCA spectra of GRS~1915+105 obtained during the low and high states
are compared with high-count states (On-states) and low-count states (Off-states) 
spectra during the irregular bursts observed on (a) 1997 June 18,
(b) 1997 July 10, and (c) 1997 July 12. Histograms show fitted models. 
Because of softening of hard states and hardening of soft states, the pivoting occurs 
at a higher energy.} 
\end{verse}
\end{figure}

If the wind is periodically cooled as described above, then, one would expect that
the duration of the low-count state, during which the sonic sphere is being 
filled in with outflowing matter, be correlated with the QPO frequency. After all,
location of the sonic sphere is proportional to the shock location~\cite{ijp98a,caa99}
and QPO frequency is also related to the shock location (eq. 1). A careful analysis
reveals~\cite{cm00} that the duration is inversely 
proportional to the square of the QPO frequency. Fig. 8 shows a log-log plot of these
quantities (adapted from ~\cite{cm00}) and the agreement with this prediction is excellent.

\begin{figure}
\vbox{
\vskip 0.0cm
\hskip 2.0cm
\centerline{
\psfig{figure=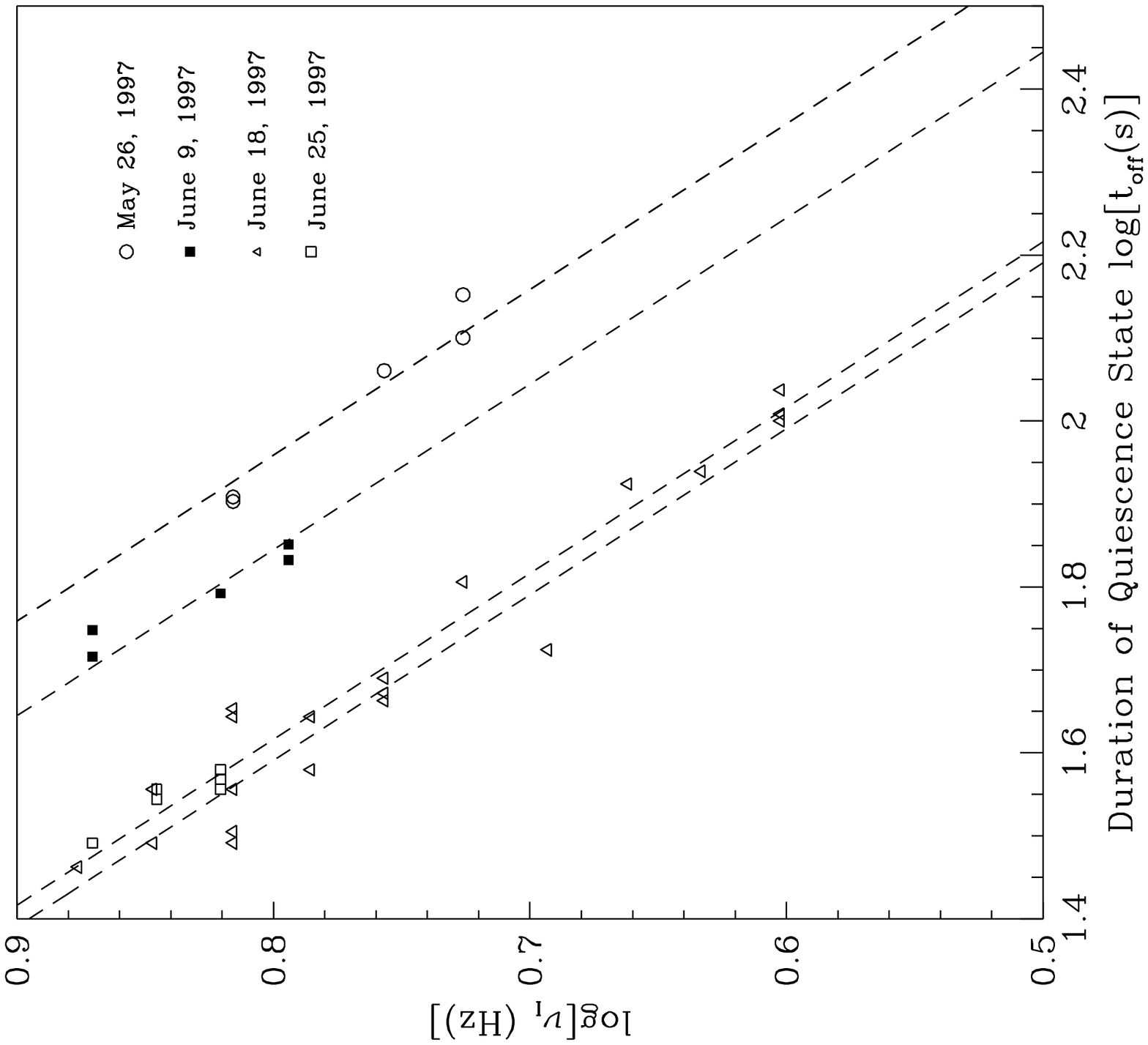,height=7truecm,width=7truecm,angle=90}}}
\begin{verse}
\vspace{0.0cm}
\noindent{\small{\bf Fig. 8:} Correlation between the duration of the low-count state
and the QPO frequency for the black hole candidate GRS 1915+105 plotted in log-log scale~\cite{cm00}.
}
\end{verse}
\end{figure}

\section*{Triumph of the advective disk paradigm}

We conclude that when the inner boundary condition on the horizon is 
considered self-consistently, the advective flow solutions obtained for various 
flow parameters, with or without shocks,  manifest themselves through various observable
effects. Given the degree of richness of the nature of the solutions it would be unwise to
simplify the study of accretion or jet processes into self-similar or other simplistic
models. A flow has to have inner sonic points, centrifugal barrier and steady or non-steady shocks
if `reasonable' parameters are assumed. Hard X-rays are generated
from the centrifugal pressure supported boundary layers (CENBOL) and not surprisingly,
behavior of outflows and rates of outflows agree qualitatively with computed values
(Eq. 2) if reasonable geometries are chosen. Even very strange behavior of 
low-count and high count-states of the oft-studied black hole candidate GRS1915+105
could be understood assuming jets are originated from the very inner region of the disk.

Since our solution is generic, results hold good for black holes of all mass. For massive black holes
time scales go up proportionately and hence switching of spectral states or
variabilities could not be studied as simply as in galactic black holes. Recent direct observations
of the radio jets in the supermassive black hole candidate M87 suggests~\cite{jbl99}
that the width of the jet could be a few tens of Schwarzschild radii near the origin. Multiwavelength
correlation in outflows of GRS1915+105~\cite{f99,d00,rm99} suggest that
the same region which produces Comptonising photons also produce jets. These observations
are sure signs that advective flow solutions which we have been proposing for over a
decade could be the basis of all future studies.

This work is partly supported by DST project Analytical and Numerical Studies of Astrophysical Flows
Around Compact Objects.

\end{document}